\documentclass[a4paper,11pt]{amsart}

\usepackage{amsmath,amssymb,amsfonts,latexsym}
\usepackage[english]{babel}
\usepackage[dvipsnames]{xcolor}
\usepackage{multirow}
\usepackage{tikz}
\usepackage{colortbl}
\usepackage{slashbox}
\usepackage{arydshln}
\usetikzlibrary{matrix,shapes,arrows,positioning}
\definecolor{myblue}{RGB}{209,221,243}
\definecolor{myblu}{RGB}{110,148,216}
\definecolor{myre}{RGB}{255,188,188}
\definecolor{li}{gray}{0.97}
\definecolor{lt}{gray}{0.7}
\usepackage{tabu}
\usepackage{pifont}
\newcommand{\cmark}{\ding{51}}%
\newcommand{\xmark}{\ding{55}}%
\usepackage{hhline}
\usepackage{algorithm}
\usepackage[noend]{algpseudocode}
\makeatletter
\def\BState{\State\hskip-\ALG@thistlm}
\usetikzlibrary{shadows}
\usepackage{booktabs}
\newcommand{\capt }[2]{\vspace{1em} \\ T{\footnotesize ABLE} #1. #2}

\newcommand{\rul}[1]{\textcolor{red}{\underline{\textcolor{black}{#1}}}}

\renewcommand{\algorithmicrequire}{\textbf{Input: }}
\renewcommand{\algorithmicensure}{\textbf{Output: }}

\begin{document}

\title[Cryptanalysis of Merkle-Hellman cipher using PGA ]{cryptanalysis of Merkle-Hellman cipher using Parallel Genetic Algorithm}

\author{KANTOUR Nedjmeddine}
\email{nkantour@usthb.dz}

\author{BOUROUBI Sadek }
\email{sbouroubi@usthb.dz}
\address{USTHB, Faculty of Mathematics,
P.B. 32 El-Alia, 16111, Bab Ezzouar, Algiers, Algeria.}


\begin{abstract}
In 1976, \textit{Whitfield Diffie} and \textit{Martin Hellman} introduced the
public key cryptography or asymmetric cryptography standards. Two years later, an asymmetric cryptosystem was published by \textit{Ralph Merkle} and \textit{Martin Hellman} called $ \mathcal{MH} $, based on a variant of knapsack problem known as the subset-sum problem which is proven to be \textit{$ \mathbb{\textit{NP}} $-hard}. Furthermore, over the last four decades, Metaheuristics have achieved a remarkable progress in solving \textit{$ \mathbb{\textit{NP}} $-hard} optimization problems. However, the conception of these methods raises several challenges, mainly the adaptation and the parameters setting. In this paper, we propose a Parallel Genetic Algorithm (PGA) adapted to explore effectively the search space of considerable size in order to break the $ \mathcal{MH} $ cipher. Experimental study is included, showing the performance of the proposed attacking scheme and finally concluding with a comparison with the $LLL$ algorithm attack.   \\

\noindent\textsc{2010 Mathematics Subject Classification.} 94A60, 68T20.

\vspace{2mm}

\noindent\textsc{Keywords and phrases.} Cryptanalysis, Merkle-Hellman Cryptosystem, Knapsack Problem, Genetic Algorithm, $LLL$ Algorithm.\vspace{-2em}

\end{abstract}

\thanks{This work was supported by L’IFORCE Laboratory of USTHB University}


\maketitle

\section { Introduction}

The Merkle-Hellman Cipher has been a subject to several attacks (see \cite{6}, \cite{7}, \cite{8}, \cite{9} and \cite{19}) namely the one proposed by A. Shamir in 1982 revealing some flaws in the reconstruction of the super-increasing sequence from the trapdoor sequence \cite{6}, nevertheless, this attack is only suitable under certain circumstances. Furthermore, there are also heuristic based attacks (Genetic Algorithm, $LLL$ Algorithm, \dots), the common factor among these proposed heuristic approaches is that they all proceed one block at a time attack. Thus, since the aim is to solve multiple knapsack problems renowned \textit{$ \mathbb{\textit{NP}} $-hard}, these approaches can be too expensive, to this point, according to our state of the art, no experimental results has been published demonstrating the performances and limits of these latter. In this paper we present a parallel heuristic approach using a Genetic Algorithm to decrypt a ciphered information in full.

\section {Merkle-Hellman cipher}

The Merkle-Hellman Cipher is one of the first asymmetric cryptosystems  proposed in 1978 
based on the knapsack problem, although this problem is proven to be \textit{$ \mathbb{\textit{NP}} $-hard} \cite{3}. However, there exists some instances where it can be solved in a linear time, as in the case of super-increasing sequence. Therefore, Merkle and Hellman proposed an arithmetical transformation ensuring the transition form a trivial knapsack based on super-increasing sequence to a hard one, identified as a trapdoor sequence. In breif, this cryptosystem uses a trivial knapsack problem which serves as a private key, and then transforms it to a general knapsack problem serving as a public key \cite{1}. These keys are generated thusly:\\
The private key consists of a super-increasing sequence $ \{a_{1} , a_{2} , \dots , a_{n} \}$, an integer $ m $ with $
{ m > \sum_{i=1}^{n} a_{i}}
 $, an integer $ w \in \{ 1, \dots ,m-1\} $ which is prime with $ m $, and a permutation $ \delta $ of the set $ \{ 1, \dots ,n\} $; hence, the public key is deduced from the private key by calculating $ b_{i}=a_{\delta(i)}w\;[m] $, for each $ i \in \{ 1, \dots ,n\}$.
Let  $ M = m_{1}m_{2}\dots m_{k}$ be a plain information  written in binary (ASCII code), where  $ m_{i}=x^{i}_{{\tiny 1}}x^{i}_{{\tiny 2}}\dots x^{i}_{{\tiny n}},\  x^{i}_{{\tiny j}} \in \{ 0,1 \} ,\; \forall i \in \{ 1, \dots , k\}$, and let $\mathcal{C}=( c_{1}, c_{2}, \dots, c_{k}) $ be the correspondent ciphered information, with $ { c_{i} = \sum_{j=1}^{n} b_{j}x^{i}_{j}},\; \forall i \in \{ 1, \dots , k\};$ to decrypt $ C $ we proceed as follows, calculating $ D_{i} = w^{-1} c_{i}\,[m] ,\; \forall i \in \{ 1, \dots , k\},$
secondly, we solve the trivial knapsack $  D_{i} = \sum_{j=1}^{n} a_{j}e^{i}_{j} ,\; $ $\forall i \in \{ 1, \dots , k\}$, finaly, deducing $ x^{i}_{j} $ from $ x^{i}_{j}= e^{i}_{\delta (j)},\; \forall i \in \{ 1, \dots , k\},\; \forall j \in \{ 1, \dots , n\} $.

\section{Cryptanalytic process}

We want to carry out an attack on a ciphered information $ C=( c_{1}, c_{2}, \dots, c_{k}) $ with the aim of finding the correspondent clear information, this amounts to uncovering the information $ M^{\star} = m^{\star}_{1}m^{\star}_{2}\dots m^{\star}_{k},\, $ $ m^{\star}_{i}=x^{{\star}_i}_{{\tiny 1}}x^{{\star}_i}_{{\tiny 2}}\dots x^{{\star}_i}_{{\tiny n}}, \; x^{{\star}_i}_{{\tiny j}} \in \{ 0,1 \} ,\; \forall i \in \{ 1, \dots , k\} $, ensuring~ :
\vspace{0.5em}
   \begin{center}
   $ {\displaystyle c_{i}-\sum_{j=1}^{n} b_{j}\,x^{{\star}_i}_{j}=0},\; \forall i \in \{ 1, \dots , k\}, $
    \end{center}
\vspace{0.5em}
where $ \{b_{1} , b_{2} , \dots , b_{n} \}$ is the public key.\vspace{0.5em}\\
Therefore, we associate the mathematical model $(\mathcal{P})$ where the plain information $M^{\star}$ represent its optimal solution, this model is well known as the Multiple Knapsack Problem.

\begin{center}
$(\mathcal{P})
\left\{
\begin{array}{l}
\vspace{-0.5cm}\\
{
\displaystyle M\textit{in\,}\mathcal{(Z)}=\sum_{i=1}^{k}\sum_{j=1}^{n} b_{j}x^{i}_{j},}\vspace{0.2em}\\
\displaystyle \sum_{j=1}^{n} b_{j}\,x^{i}_{j} \geq c_i , \; \forall i\in \{1, \dots , k\},\vspace{0.2em}\\
{\displaystyle x^{i}_{j} \in \{0,1\} ,\: \forall j\in \{1, \dots , n\}. }\vspace{2mm}
\end{array}
\right.
 $ \vspace{0.5em}
\end{center}

Intending to construct a flexible mathematical model, and considering the fact that only the optimal solution breaks the ciphered information (the best or nothing situation), for each bloc $ i $ we associate a cryptanalysis model $(\mathcal{P}_i)$ to each ciphered block $i$ where its optimal solution is $m^{\star}_{i}$, moreover, we can rely on the following model while it retains the same objective values: \vspace{0.5em}
\begin{center}
$(\mathcal{P}_{i})
\left\{
\begin{array}{l}
\vspace{-0.5cm}\\
{
\displaystyle M\textit{in}\mathcal{(Z)}=|c_{i}-\sum_{j=1}^{n} b_{j}x^{i}_{j}|,}\vspace{0.3em}\\

{\displaystyle x^{i}_{j} \in \{0,1\} ,\: \forall j\in \{1, \dots, n\}. }\vspace{2mm}
\end{array}
\right.
 $ \vspace{0.5em}
\end{center}

To find the original information we are required to solve numerous \textit{$ \mathbb{\textit{NP}} $-hard} problems $ (\mathcal{P}_{i})_{1<i\leq k} $, thus, we  propose an attacking scheme that mainly uses  multiple genetic algorithms in a parallel way, in which they are enhanced with distinctive search and communication strategies.

\section {Genetic Algorithm}
Genetic algorithms have been developed by John Holland in 1975, presented in his book \textit{Adaptation in Natural and Artificial Systems} \cite{13}, which roughly, are optimization algorithms inspired from natural evolution mechanisms and genetic science, the main proposed idea was to combine the principle "survival of the fittest" among string structures with a deliberated random information exchange \cite{10}; initiated with a set of  chromosomes (i.e potential solutions) called initial population, this algorithm uses "natural selection" along with genetics-inspired operators (crossover, mutation and selection) to move from one population to a new population \cite{11}, hence, reducing the search procedure among exponential-ordered search space to a collection of candidate solutions incrementally adapted, regarding the adaptation measurement, a fitness function is defined to evaluate chromosomes according to their performance against a given problem. In the rest of this section we present an adaptation of the genetic algorithm with slight modifications to solve the knapsack problem $(\mathcal{P}_i)$ defined above. Starting with the representation of the chromosomes $m_i$ (i.e solutions of the problem $(\mathcal{P}_i)$), we have chosen to use a natural representation so as a chromosome is a binary string, thus, $m_i= x^{_i}_{{\tiny 1}}x^{_i}_{{\tiny 2}}\dots x^{_i}_{{\tiny n}}, \; x^{_i}_{{\tiny j}} \in \{ 0,1 \}$, where $n$ is the public key length. To initiate the algorithm we need to generate an initial population, for that, we used a greedy  heuristic that generates randomly a large set of chromosomes and then retains a given number of best fitted ones, so as to provide a set of chromosomes of "acceptable" quality; when entring the algorithm, we randomly select parents form the current population  according to a mating probability (called also crossover probability) $P_c$ , the selected chromosomes are in charge of producing the next population. For that purpose, we conduct three operations~: crossover, mutation and selection. The crossover operator is defined in genetic algorithms as an analogy of the mecanism that allows the reproduction of chromosomes in nature, where the produced chromosomes partially heretates caracteristics from its parents with copying and recombinnig their genes in a delaberated way. For the mating process we used three crossover operators proceeding for each pair, first, we use a very classical operator, choosing a cutting point, say $c \in \{2,\dots,n-1 \}$, here the mating process consists of swapping bits $c+1$ to $r$ of the first parent $P_1$ with bits $c+1$ to $r$ of the second parent $P_2$ \cite{7}, hence, new chromosomes $C_1$ and $C_2$ are created.\vspace{-6mm}\\
\hspace{-0.em}
\begin{figure}[h!]
\centering
\begin{tabular}{c|c|c|c|c|c!{\color{red}\vline width 3pt}c|c|c|c|c|c|c|c|c|c|c|c|c}

$P_1$ = &1&1&1&0&1&\textbf{1}&\textbf{0}&\textbf{0}& $\longrightarrow$ $C_1$ = &1&1&1&0&1&\textbf{1}&\textbf{1}&\textbf{1}\\ \cline{2-6}
\end{tabular}\vspace{0.6em}\\ \hspace{-5.5cm}
\begin{tabular}{ccccccccc}

& & & \;swap & $\updownarrow$& & & & \\
\end{tabular}\vspace{0.1em}\\
\begin{tabular}{c!{\color{black}\vline width 0.5pt}c|c|c|c|c!{\color{red}\vline width 3pt}c|c|c|c|c|c|c|c|c|c|c|c|c}
\cline{2-6} \vspace{-1.15em} \\
$P_2$ = &0&1&1&0&0&\textbf{1}&\textbf{1}&\textbf{1}& $\longrightarrow$ $C_2$ = &0&1&1&0&0&\textbf{1}&\textbf{0}&\textbf{0}\\
\end{tabular}

\caption{One cutting point crossover operator.}
\end{figure}

The second operator is similar to the first one but here we use tow cutting points, and new chromosomes are generated swapping the middle parts of the parents, as it shown in Figure 2.\vspace{-2mm} \\
\begin{figure}[h]
\centering
\begin{tabular}{c|c|c!{\color{red}\vline width 3pt}c|c|c|c!{\color{red}\vline width 3pt}c|c|c|c|c|c|c|c|c|c|c|c}

$P_1$ = &1&1&\textbf{1}&\textbf{0}&\textbf{1}&\textbf{1}&0&0& $\longrightarrow$ $C_1$ = &1&1&\textbf{1}&\textbf{0}&\textbf{0}&\textbf{1}&0&0\\ \cline{4-7}
\end{tabular}\vspace{0.6em} \\ \hspace{-8.3cm}
\begin{tabular}{ccccccccc}

& & & & & & swap & $\updownarrow$ & \\
\end{tabular} \vspace{0.3em}\\
\centering
\begin{tabular}{c!{\color{black}\vline width 0.5pt}c|c!{\color{red}\vline width 3pt}c|c|c|c!{\color{red}\vline width 3pt}c|c|c|c|c|c|c|c|c|c|c|c}
\cline{4-7} \vspace{-1.15em} \\
$P_2$ = &0&1&\textbf{1}&\textbf{0}&\textbf{0}&\textbf{1}&1&1& $\longrightarrow$ $C_2$ = &0&1&\textbf{1}&\textbf{0}&\textbf{1}&\textbf{1}&1&1\\
\end{tabular}
\caption{Two cutting point crossover operator.}
\end{figure}\vspace{-2mm} \\
The third operator constructs one child chromosome $C$ by alternately choosing random genes from the parents (see Figure 3), this operator is known as the uniform crossover operator.
\setlength\doublerulesep{1pt}
\doublerulesepcolor{red}
\begin{figure}[h]
\centering
\begin{tabular}{c|c|c|c|c|c|c|c|c|}
$P_1$ = &\textbf{\rul{1}}&\textbf{\rul{ 1}}&1&0&\textbf{\rul{ 1}}&1&0&\textbf{\rul{0}} \\
\multicolumn{9}{c}{}\\
$P_2$ = &0&1&\textbf{\rul{ 1}}&\textbf{\rul{0}}&0&\textbf{\rul{ 1}}&\textbf{\rul{1}}&1
\end{tabular} \hspace{-1em}
$ \left.
\begin{array}{c}
\vspace{.2em}\\
\vspace{.2em}\\
\end{array}
\right\rbrace
$\hspace{-0.9em}
\begin{tabular}{cc|c|c|c|c|c|c|c|c|}
\vspace{-0.1em}\\
& \hspace{-.2cm}${\footnotesize \rightarrow\quad} C$ = &1&1&1&0&1&1&1&0 \\
 \\
\end{tabular}
\caption{Uniform crossover operator.}
\end{figure}\vspace{-2mm}\\
The newly produced chromosomes are added to the current population, where at this point, we perform mutation according to a probability $p_m$, here, its  value is usually chosen less than $0.1$. Furthermore, the used operator consists of replacing a block of genes by its complementary, this latter are generally of length one.
\begin{figure}[H]
\centering
\begin{tabular}{c|c|!{\color{red}\vline width 0.5pt}c!{\color{red}\vline width 0.5pt}|c|c|c|c|c|c|c|c|!{\color{red}\vline width 0.5pt}c!{\color{red}\vline width 0.5pt}|c|c|c|c|c|c|}
\arrayrulecolor{red}\cline{3-3} \cline{12-12} \arrayrulecolor{black}
$P_1$ = &1&1&1&0&1&1&0&0& ${\footnotesize \rightarrow}$ $P^{'}_1$ = &1&0&1&0&1&1&0&0\\
\arrayrulecolor{red}\cline{3-3}\cline{12-12} \arrayrulecolor{black}
\end{tabular}
\vspace{1em}
\caption{Mutation operator.}\vspace{-.3cm}
\end{figure}
Aspiring to improve the fitness of chromosomes in the current population (including the offsprings produced in the previous steps) and accelerate the search process before launching the selection operator, we apply a low-cost improving heuristic that essentially withdraws a given number of chromosomes and then tempts to apply on each drawn chromosome corrections on its genes in order to extract form it a better fitted chromosome, by bringing chromosomes to a local optima, in case of a positive result, the produced chromosomes will be added to the current population replacing the existing parents.
The following step is to select the chromosomes to maintain for the next generation, in which they will serve as parents. Hence, this operator can be critical, since here,  we select the chromosomes to survive and potentially spread their genes in forthcoming generations. In contrast, we consequently eliminate chromosomes qualified by the chosen operator as unadapted. We used two  randomly chosen operators according to a probability $p_s$. To begin with, we employed the elitist selection, that consists to retain the best fitted chromosomes for the next generation, this operator prevent the loss of "good" genes, in contrast, it can cause a premature convergence. The other used operator is known as the roulette wheel selection, this latter consists to associate for each chromosomes $i$ a probability  of selection $p_i$ according to its fitness value $f_i$. Moreover, we allocate for each chromosome $i$ a proportion of length $p_i$ in the segment $[0,1]$ (an angle of $2\pi p_i$ in case of circular representation), for that we use the Algorithm 1 below:

\begin{algorithm}
\caption{}
\begin{flushleft}
\algorithmicrequire Fitness values   $f_1,\dots,f_n$ \\
\algorithmicensure Selection probabilities  $p_1,\dots,p_n$\\ \vspace{-0.29cm}
\rule{\linewidth}{0.5pt}\vspace{-0.1cm}
\end{flushleft}\begin{algorithmic}[1]
\State Calculate $f_{max}:=\, \underset{1\leq i\leq n} \max\{\,f_i\,\}\,;$
\State For each $i \in \{1, \dots , n \},\; f_i :=f_{max}-f_i+1;$\vspace{0.3cm}
\State For each \vspace{-0.59cm}$$\hspace{-3.cm} i \in \{1, \dots , n \},\; p_i :=\frac{f_i}{\sum^{n}_{j=0} f_j};$$ \vspace{-0.3cm}
\State \textbf{Return} $p_1,\dots,p_n.$
\end{algorithmic}
\end{algorithm}
Let $P=\{I_1,I_2,\dots,I_7\}$ a population (a set of chromosomes as defined privously), and respectively their fitness evaluations $f_1=100,\,f_2=50,\,f_3=60,\,f_4=160,\,f_5=30,\,f_6=40,\,f_7=10$, by applying the algorithm above we obtained respectively the proportions  $p_1=0.08,\,p_2= 0.115,\,p_3=0.085,\,p_4=0.15,\,p_5=0.26,\,p_6=0.16,\,p_7=0.15$; with which, we constructed the wheel in Figure 5; after that, one "random" positions in the wheel is generated at a time to pinpoint a chromosome to be selected. For instance, the value $0.55$ (and any value between $0.43$ and $0.69$) would select the chromosome $I_5$. Furthermore, there are many variants of this operator, we name the stochastic universal selection (SUS) introduced by James Baker \cite{4}, this operator selects simultaneously multiple chromosomes by choosing a given number of equally spaced values, as it is shown in Figure 5, the spinners indicates to select three chromosomes $I_3, I_5, I_7$. This operator has an advantage over the roulette wheel selection, in case we have a chromosome that dominates the population (in terms of its proportion in the wheel), in which, the latter can be selected excessively.
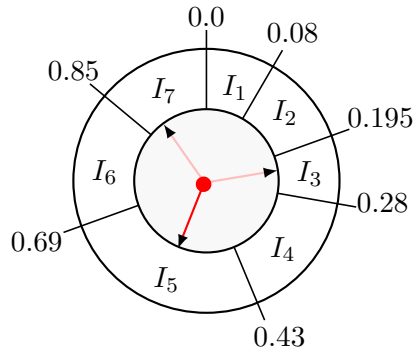
\begin{figure}[h]
\centering
\begin{tikzpicture}[>=latex,font=\sffamily,semithick,scale=2.5]
\draw [thick] (0,0) circle (.7);

\node (0,0) {\rotatebox{68}{\rotatebox{-68}{$I_{5}$} \, \, \, \, \, \, \, \, \, \, \, \, \, \, }} ;

\node (0,0) {\rotatebox{-5}{\rotatebox{5}{$I_{ 6\,}$} \, \, \, \, \, \, \, \, \, \, \, \, \, }} ;

\node (0,0) {\rotatebox{-66}{\rotatebox{66}{$I_{7\,}$} \, \, \, \, \, \, \, \, \, \, \, \, \, }} ;

\node (0,0) {\rotatebox{-110}{\rotatebox{110}{$I_{1\, \, \,}$} \, \, \, \, \, \, \, \, \, \, \, \, \, }} ;

\node (0,0) {\rotatebox{140}{\rotatebox{-140}{$I_{4\, \, \,}$} \, \, \, \, \, \, \, \, \, \, \, \, \, \, }} ;

\node (0,0) {\rotatebox{184}{\rotatebox{-184}{$I_{3\, \,}$} \, \, \, \, \, \, \, \, \, \, \, \, \, \, }} ;

\node (0,0) {\rotatebox{220}{\rotatebox{-220}{$I_{2\, \,}$} \, \, \, \, \, \, \, \, \, \, \, \, \, \, }} ;

\node (0,0) {\rotatebox{-90}{\rotatebox{90}{$0.0$} \, \, \, \, \, \, \, \, \, \, \, \, \, \,\, \,\, \,\, \,\, \,\, \,\, \, }} ;
\node (0,0) {\rotatebox{240}{\rotatebox{-240}{$0.08$} \,\, \ \, \, \, \, \, \, \, \, \, \, \, \, \,\, \,\, \,\, \,\, \,\, \,\, \, }} ;
\node (0,0) {\rotatebox{200}{\rotatebox{-200}{$0.195$} \, \, \, \ \, \, \, \, \, \, \, \, \, \, \, \, \,\, \,\, \,\, \,\, \,\, \,\, \, }} ;
\node (0,0) {\rotatebox{173}{\rotatebox{-173}{$0.28$} \, \, \ \, \, \, \, \, \, \, \, \, \, \, \, \,\, \,\, \,\, \,\, \,\, \,\, \, }} ;
\node (0,0) {\rotatebox{115}{\rotatebox{-115}{$0.43$} \,\, \ \, \, \, \, \, \, \, \, \, \, \, \, \,\, \,\, \,\, \,\, \,\, \,\, \, }} ;
\node (0,0) {\rotatebox{20}{\rotatebox{-20}{$0.69$} \, \, \,\ \, \, \, \, \, \, \, \, \, \, \, \, \,\, \,\, \,\, \,\, \,\, \,\, \, }} ;
\node (0,0) {\rotatebox{320}{\rotatebox{-320}{$0.85$} \,\, \ \, \, \, \, \, \, \, \, \, \, \, \, \,\, \,\, \,\, \,\, \,\, \,\, \, }} ;
\foreach \angle in {90,60,20,-10,-67.5,200,140}
\draw   (0,0) --(\angle:.8cm) ++  (\angle:0.2cm) ;

\node [circle,thick,fill=li,draw=black,align=center,minimum size=1.9cm] at (0,0) {};

\draw [->,thick,draw=myre] (-0.01,-0.01) -- (-0.225,0.3);
\draw [->,thick,draw=myre] (-0.01,-0.01) -- (0.375,0.055);
\draw [*->,thick,draw=red] (0,0.02) -- (-0.15,-0.36);
\end{tikzpicture}
\vspace{-.5cm}
\caption{Roulette wheel selection.}
\end{figure}\newpage
As has been mentioned earlier, the selection operator is chosen according to a given probability $p_s$. In practice, we recommend the values in $[0.25,\,0.3]$ as to set an equilibrium between a further exploitation of the current best fitted chromosomes and diversification illustrated as the exploitation of potential "partially fitted" chromosomes, that is, by altering between elitist and roulette wheel selection operators.
\section {Cryptanalytsis scheme}
Commonly, parallel computing is considered as a means to improve the performance of programs that have high computational cost (reduces the execution time), as in the process of solving a  \textit{NP-hard} problems, in which, recourse is often made to metaheuristics. One of the most studied yet effective metaheuristics is the Genetic Algorithm (GA), the parallelization of this latter is a technic that is used to deal with large instances, aiming not only to reduce the computation time but also to improve the quality of solutions, inducing higher effectiveness compared to sequential GAs, especially, by the arrival of cooperative multi-search models. In this study, the intuitive approach of parallelization is to carry out a search (attack) on each block separately; however, the employed approach is a distributed attack preformed on each block and enhanced with a communication strategy among parallel processing elements (PEs) known as the migration in PGA,  where the latter is established in order to provide a cooperation among parallel PEs. However, since search elements have different objectives, the utilization of the migration operator, which we will elaborate later, is justified as an exploitation of the sensibility of the trapdoor sequence. Let $m_i,m_j \in {\{0,1\}}^{n}$ tow blocks plain information of length $2n$ and $\Phi_i$ is a function that operates on the elements of a block $i$, where $\Phi_i (m)= c_i\,(\sum_{j} b_j x_j )^{-1}, \, m \in {\{0,1\}}^{n}$. Since in practice, the obtained ciphered information $c_i$ can be colossal for large instances, for instance, a one block ciphered information of length $64$ usually exceeds $10^{20}$, hence, the function $\Phi$ is used to reduce the value of ciphered informations to an approximated value in $[0,1]$. In order to study the sensibility of the public key, we construct a subset of chromosomes from $N(m_i)$ the neighborhood of $m_i$ the optimal solution for block $i$ and we compare their fitness in different environments (i.e. fitness in each block attack); in practice we used a set $ J \subset E = \{ m \, | \, 0 < d(m, m_i)\leq 3 \} \subset N(m_i)$, with $d(x\,,y), \, x,y \in {\{0,1\}}^{n} $ is the Hamming distance between $x$ and $y$, where the set $J$ is obtained by randomly applying one to three bit modification on $m_i$; we observe that: $$\exists \, m \in N(m_i),\, |\Phi_i(m)-\Phi_j(m_j)|<|\Phi_i(m)-\Phi_i(m_i)|,$$ in other terms, a slight modification in a chromosome can change the environment in which it adapts, confirming that we can extract a near optimal solution for a certain block attack from other block's current population, moreover, it is quite possible even for the most adapted chromosomes to induce adapted solutions for other environments.

On this basis, we can circumstantiate the importance of establishing a communication link among GAs, illustrated as a migration operator that consists of sending chromosomes that verify the migration condition from a population to another, through connections along all GAs. For each block $j \in \{ 1, \dots, k\}$ the attacking process is summarized in the Algorithm~2.\vspace{1em}
\begin{algorithm}
{\fontsize{12pt}{12pt}\selectfont
\caption{Parallel Genetic Algorithm  (block $j$)}\label{alg:HI}
\begin{algorithmic}[l]
\State \vspace{0.2em}
\State \hspace{-0.5cm}\textbf{Begin} \vspace{0.2em}
\State Initialize the Number of Generations $ nb := 0 $;
\vspace{0.1cm}
\State Generate the initial population $P_{j}^{\,0}$;
\vspace{0.1cm}
\State Evaluate chromosomes in $P_{j}^{\,0}$;
\vspace{0.1cm}
\While{(Plain information $m_j$ not found)\vspace{0.1cm}}
        \State Select parents form $P_{j}^{{\scriptsize\,nb}}$ and Apply the crossover operator \\ \hspace{4.5mm} according to $p_c$;\vspace{0.1cm}
        \State Apply the mutation operator according to $p_m$;\vspace{0.1cm}
        \State Apply an improving heuristic;\vspace{0.1cm}
        \State Evaluate the chromosomes in $P_{j}^{\,nb}$;\vspace{0.1cm}
        \If{(Migration condition is satisfied)\vspace{0.1cm}}
            \State Send the concerned chromosomes to their new population;\vspace{0.1cm}
            \State \textbf{endif}\vspace{0.1cm}
        \EndIf
        \If{(Receiving chromosomes)\vspace{0.1cm}}
            \State Remove the 'worst' chromosomes and replace them with the \\ \hspace{1.07cm} received ones;\vspace{0.1cm}
            \State \textbf{endif}\vspace{0.1cm}
         \EndIf
         \State Increment $nb$;
         \State \hspace{-4mm}\textbf{endwhile}\vspace{0.1cm}
      \EndWhile

      \State \vspace{.7mm} Apply a selection operator chosen according to $p_s$;

\State \hspace{-0.5cm}\textbf{End} \vspace{0.4em}
\end{algorithmic}
}
\end{algorithm}\vspace{2mm}
\\
Regarding the migration operator we have chosen a state dependent operator that operates as follows: we evaluate each chromosome in $P_i$ the current population of the block $i$ according to their fitness in the rest of the blocks $\{1,\dots,i-1,i+1,\dots,k\}$ and we compare each of them with the best fitted chromosome in the current populations $\{P_1,\dots,P_{i-1},P_{i+1},\dots,P_k\}$ (compared with the best solution in the current population which is not necessarily the best found solution), in other words, let  $m \in {\{0,1\}}^{n}$ and $f_i(m)=|c_{i}-\sum_{j=1}^{n} b_{j}x^{i}_{j}|$ the fitness function associated to the $i^{th}$ block; we verify for each bloc $i \in \{1,\dots,k\} $ if there exists a chromosome $m'\in P_i$ that satisfies :
$$\forall\, m \in P_j,\, f_j(m')<f_j(m), \, j\in\{1,\dots,i-1,i+1,\dots,k\},$$
then the chromosome $m'$ is sent to $P_j$ the current population of the block $j$, in which it will replace the 'worst' chromosome in the recipient population. However, regarding that the blocks has different objectives, the migration operator is applied before the selection operator in order to exploit the unfitted chromosomes risking to be removed in the process of selection. \\In Figure 6, we put forward a summation of the attacking scheme presented in the previous sections.
\vspace{2mm}
\tikzset{
    arr/.style={->,blue,very thick},
    lbl/.style={draw,blue,very thick},
    shadowed/.style={preaction={transform canvas={shift={(0pt,-2pt)}},draw=lightgray,fill=lightgray,thick}},
    shadowe/.style={preaction={transform canvas={shift={(2pt,-2.3pt)}},draw=lightgray,fill=gray,thick}},
  }

\tikzstyle{dia} = [diamond, draw, fill=white,
    text width=7em, text centered,shadowed, node distance=1mm, rounded corners=.1cm,inner sep=0pt,aspect=2.5]

    \tikzstyle{bloc} = [rectangle, draw, minimum width=3cm,
 draw, fill=white,
    text width=5em, text centered,node distance=2.3cm and 1cm,shadowed,rounded corners=.15cm, minimum height=2em]

 \tikzstyle{blo} = [trapezium, draw, minimum width=3.3cm,
trapezium left angle=120, trapezium right angle=60, draw, fill=white,
    text width=5em, shadowed,text centered,node distance=1.cm and 1.3cm, rounded corners=.1cm,minimum height=2.em]

\tikzstyle{block} = [trapezium, draw, minimum width=3cm,
trapezium left angle=0, trapezium right angle=0, draw,fill=white,
    text width=5em,shadowed, text centered,node distance=2.3cm and 1cm,rounded corners=.07cm, minimum height=1em]

\tikzstyle{line} = [draw, -latex']
\tikzstyle{cloud} = [draw, ellipse,fill=red!20, node distance=1cm,
    minimum height=3em]
    \tikzstyle{line1} =
    [
        draw,,red
     , -latex'
     ]
     \tikzstyle{line2} =
    [
        draw,,blue
     , -latex'
     ]
    \tikzstyle{line4} =
    [
        draw,black
     , -latex'
     ]
   \begin{figure}[h!]

    \begin{center}
     \begin{tikzpicture}[scale=0.8, transform shape]

		\node (A)
		[	
		     fill = lt, rectangle, text centered,rounded corners = 0.5pt,
		    shadowe,draw=black, fill=li, opacity=1,
		     text badly ragged ,inner sep=3pt
		]
	{
	\hspace{1mm}
\begin{tikzpicture}

  \matrix (mat) [matrix , column sep=16mm, row sep=0.8cm]
{
    \node [bloc,drop shadow,line width=0.5mm] (Bloc 1) {Bloc $ 1 $ : $ c_1 $}; &
    \node [bloc,line width=0.5mm] (Bloc 2) {Bloc 2 : $ c_2 $}; &
    \node [bloc,line width=0.5mm] (Bloc 4) {bloc $ k $ : $ c_k $};\\
    \node [block] (heur1) {Initialize~$P_{1}$};&
    \node [block] (heur2) {Initialize~$P_{2}$}; &
    \node [block] (heur4) {Initialize~$P_{k}$};\\
    \node [block] (B1) {$GA_1$}; &
    \node [block] (B2) {$GA_2$}; &
    \node [block] (B4) {$GA_k$};\\
    \node [blo] (G1) {Generation~$ i_1 $};&
    \node [blo] (G2) {Generation~$ i_2 $};&
    \node [blo] (G4) {Generation~$ i_k $};\\
    \node [dia] (D1) {Decrypt($ c_1 $)};&
    \node [dia] (D2) {Decrypt($ c_2 $)}; &
    \node [dia] (D4) {Decrypt($ c_k $)};\\
    \node [bloc,line width=0.5mm] (I1) {Bloc$\, 1 $: $ m_1 $}; &
    \node [bloc,line width=0.5mm] (I2) {Bloc$ \,2 $: $ m_2 $}; &
    \node [bloc,line width=0.5mm] (I4) {Bloc$\, k $: $ m_k $};\\
};

\path[line,li,thick] (Bloc 2.east)    --+(0.3,0.) node [pos=0.7, right,black] { {\hspace{6mm}\LARGE \textbf{\dots}}  }-- (Bloc 4.west);
\path[line,li,thick] (I2.east)    --+(0.4,-0.3) node [pos=0.7, right,black] { {\hspace{6.7mm}\LARGE \textbf{\dots}}  }-- (I4.west);

\path[line1,thick] (Bloc 1.south)    --+(0,-0.5) -|  (heur1.north);
\path[line2,thick] (Bloc 2.south)    --+(0,-0.5) -| (heur2.north);
\path[line4,thick] (Bloc 4.south)    --+(0,-0.5) -|  (heur4.north);

\path[line1,thick] (heur1.south)    --+(0,-0.5) -| (B1.north);
\path[line2,thick] (heur2.south)    --+(0,-0.5) -|  (B2.north);
\path[line4,thick] (heur4.south)    --+(0,-0.5) -|  (B4.north);

\path[line1,thick] (B1.south)    --+(0,-0.5) node [pos=0.7, right] { $ i_1  $:=$ i_1 +1 $;  }-| (G1.north);
\path[line2,thick] (B2.south)    --+(0,-0.5) node [pos=0.7, right] { $ i_2  $:=$ i_2 +1 $;  }-| (G2.north);
\path[line4,thick] (B4.south)    --+(0,-0.5) node [pos=0.7, right] { $ i_k  $:=$ i_k +1 $;  }-| (G4.north);

\draw[->,>=latex,red] (G1.south)     -- (D1.north);
\draw[->,>=latex,densely dashed,blue] (G2.south)     --  (D1.north);
\draw[->,>=latex,densely dashed,black](G4.south)    --   (D1.north);

\draw[->,>=latex,densely dashed,red] (G1.south)     -- (D2.north);
\draw[->,>=latex,blue] (G2.south)     --  (D2.north);
\draw[->,>=latex,densely dashed,black](G4.south)    --   (D2.north);

\draw[->,>=latex,densely dashed,red] (G1.south)     --  (D4.north);
\draw[->,>=latex,densely dashed,blue] (G2.south)     --  (D4.north);
\draw[->,>=latex,black](G4.south)    --   (D4.north);

\path[line1,thick] (D1.west)    -- +(-.2,0) node [pos=0.3, above] { No } |- +(-.4,0) |- (B1.west);
\path[line2,thick] (D2.west)     -- +(-.2,0) node [pos=0.3, above] { No } |- +(-.4,0) |- (B2.west);
\path[line4,thick] (D4.west)   -- +(-.2,0) node [pos=0.3, above] { No } |- +(-.4,0) |- (B4.west);

\path[line1,thick] (D1.east)    -- +(.4,0) node [pos=0.3, above] { \ Yes } |- +(.4,0) |-  (I1.east);
\path[line2,thick] (D2.east)    -- +(.4,0) node [pos=0.3, above] { \ Yes } |- +(.4,0) |-  (I2.east);
\path[line4,thick] (D4.east)    -- +(.4,0) node [pos=0.3, above] { \ Yes } |- +(.4,0) |-  (I4.east);

\end{tikzpicture}
};
	\end{tikzpicture}
    \end{center}
    \vspace{-0.5em}
    \caption{Summation of the cryptanalysis technique.}
 \end{figure}
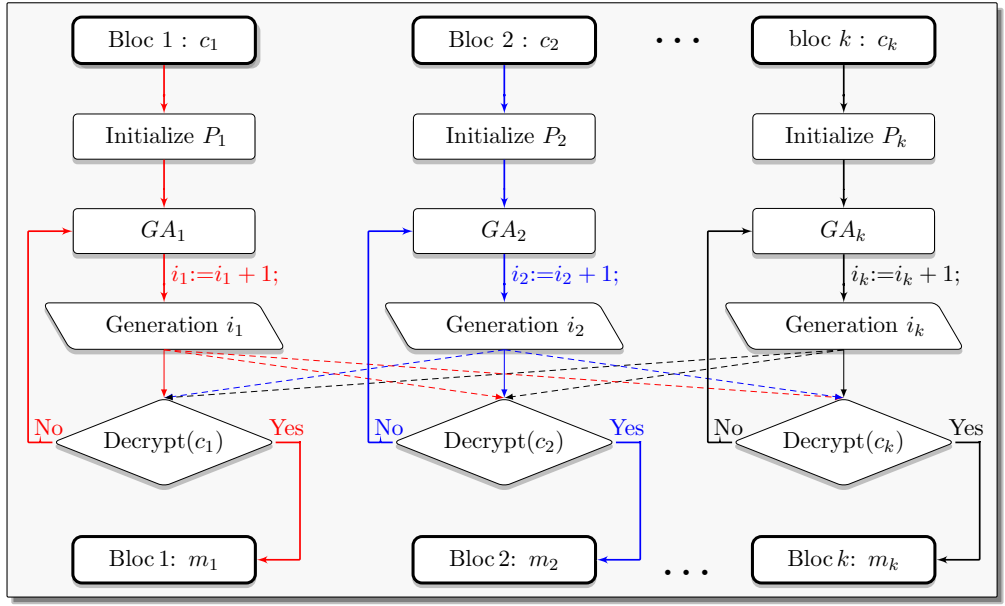

\section{Computational results}

In this section we present the numerical results we have obtained on the cryptanalysis of the Merkel-Hellman cipher using the scheme presented in the previous sections. The implementation was achieved using Java SE platform, in which the parallelization is concretized via multithreading. All the following experiments was conducted on a single machine including an Intel i7-5600U~CPU (2 cores and 2 logical cores per physical). \\The following experiments consist of studying the impact of the parameters variations on the scheme's performances, while the concerned parameters are: the population size, mutation probability $p_m$, selection probability $p_s$ and the heuristic's number of iterations. We must point that in the experimentations, all the parallel attacks (GAs) are applied with the same parameters values, anywise, the aim is to spot suitable parameters values for a given key size $n$ and study the effect of these parameters on the scheme. To achieve this, in each experimentation we tested 100 random instances for each key length, where all clear informations are coded in 8 bits printable ASCII; however, all the obtained results are adduced  according to  different key lengths $n \in \{8 \times i \, | \, 3\leq i \leq 14 \}$.\\
The following table summarizes the conducted experimentations in statistical results (minimum, maximum, median and average) according to different key lengths, showing the execution time and the number of generations, for successful attacks on a one and tow blocks ciphered text.
\begin{figure}[h!]
\centering
\begin{footnotesize}
\setlength{\tabcolsep}{0.5em}
\setlength{\arrayrulewidth}{.05em}
{\renewcommand{\arraystretch}{1.3}
\begin{tabular}{cccccc:cccc}
\hline
\multirow{2}{*}{$n$} & \multirow{2}{*}{\hspace{0.1em}$k$\hspace{0.4em}} & \multicolumn{4}{c}{Execution time (ms)} &  \multicolumn{4}{c}{Number of generations} \\
\cline{3-10}
 & & min & max & median & average & min & max & median & average\\
\hline
\multirow{2}{*}{24} &1 &11& 4309& 47& 760& 1& 1498& 14& 258\\
                    &2 & 12& 3209& 63&	293 &1& 1283& 9& 84\\
\hline
\multirow{2}{*}{32} &1 & 25& 855& 281& 367& 5& 341& 111& 129\\
                    &2 & 31& 4511& 344& 1060 & 6& 1261& 81& 308 \\
\hline
\multirow{2}{*}{40} &1 &47	&17530	&890	&1776 & 10 & 1958 & 128 & 321\\
                    &2 &141	&20481	&1384	&2922 & 41 & 4233 & 291 & 507\\
\hline
\multirow{2}{*}{48} &1 &125	&46645	&5039	&7853& 13&	6388&	487& 1040
\\
                    &2 &2261 &39368	&6766	&8711 & 333 & 7386 & 932 & 1337\\
\hline
\multirow{2}{*}{56} & 1& 149 & 49926 & 7685 & 11724 & 16 & 7829 & 702 & 1077 \\
                    & 2&515	&35942	&7462	&9752& 76 & 7519 &1186 & 1764\\
\hline
\multirow{2}{*}{64} &1 &140	&82617 &8827 &13716 &11 & 17385 & 950 & 2070\\
                    &2 &1964& 77429& 9480& 17346 &234&	11341&	739&	1738\\
\hline
\multirow{2}{*}{72} &1 &1412& 1078663& 37285& 78302 & 80& 6341& 1213& 2033\\
                    &2 &8412& 116903& 36016& 42523& 222& 9835& 1936& 3168\\
\hline
\multirow{2}{*}{80} &1 &1237& 661800& 71919& 102075 & 90& 54576& 5893& 8381\\
                    &2 &2970& 392536& 65405& 87398 & 87& 25908& 4419& 6084\\
\hline
\multirow{2}{*}{88} &1 &790& 278324& 101874 & 139909 &49& 25558& 5418& 8385\\
                    &2 &2265& 332563& 128968& 131283 &149& 27008& 10061& 10285\\
\hline
\multirow{2}{*}{96} &1 &40435& 1874332& 220279& 424788&3591& 80321& 10560& 19542\\
                    &2 &33613& 1114694& 243148& 337441&2343&65965&18254&23310\\
\hline
\multirow{2}{*}{104} &1 &102788&1770714&739343&780345&1905&58271&17467&20447\\
                    &2 &161213&2032836&786615&846839&6070&90594&23771&31351\\
\hline
\multirow{2}{*}{112} &1 &38357&2595738&1335045&1119082&1294&52353&23584&25471\\
                    &2 &27822&3695401&1356234&1308807&429&69124&24202&25427\\
\hline

\end{tabular}
}
\end{footnotesize}
\capt{1}{Summation of the experimentations.}
\end{figure}
\vspace{2em}\\
We tested the proposed scheme against variations of $k$ the number of blocks, by running three instances of size $k\in \{3,4,5\}$ of different key size (see Table~2).
\begin{figure}[h!]
\centering
\begin{footnotesize}
\setlength{\tabcolsep}{0.7em}
\setlength{\arrayrulewidth}{.05em}
{\renewcommand{\arraystretch}{1.1}
\begin{tabular}{cccc:ccc}
\cline{1-7}
  \multicolumn{1}{c}{} & \multicolumn{3}{c:}{Execution time (ms)} &  \multicolumn{3}{c}{Nb. of generations} \\
\cline{1-7}
$k$ & 3 & 4 & 5 &3 & 4 & 5 \\ \hline
{$n=$} 32 & 2326 &6895 & 9355 & 198& 340&223 \\ \hline
\textcolor[rgb]{1.00,1.00,1.00}{$n=$} 40 &18669 &7592 & 19351& 142 &361& 319\\ \hline
\textcolor[rgb]{1.00,1.00,1.00}{$n=$} 48 &18669 &7592 & 19351& 142 &361& 319\\ \hline
\textcolor[rgb]{1.00,1.00,1.00}{$n=$} 56 & 2326 &6895 & 9355 & 198& 340&223 \\ \hline
\textcolor[rgb]{1.00,1.00,1.00}{$n=$} 64 &18669 &7592 & 19351& 142 &361& 319\\ \hline
\textcolor[rgb]{1.00,1.00,1.00}{$n=$} 72 &30116 & 24249&35708 &7091 &1153 &362 \\ \hline
\textcolor[rgb]{1.00,1.00,1.00}{$n=$} 80 &27060 & 48024&77951 & 1496&2049 & 928\\ \hline
\textcolor[rgb]{1.00,1.00,1.00}{$n=$} 88 &89590 & 180732& 307455& 5486 &3016 & 2681\\ \hline
\textcolor[rgb]{1.00,1.00,1.00}{$n=$} 96 &177154 & 97566&116490 &11739 & 3567&656 \\ \hline
\end{tabular}
}
\end{footnotesize}
\capt{2}{Performance of the $PGA$ attack against the number of blocks.}
\end{figure}

The aim of the following experiment is to observe the resemblance ratio by comparing the best found solution to the plain information for either successful and unsuccessful attacks. Table 3 shows the obtained resemblance ratios for different kay lengths, where the used instances are of length $2$ to $4$ blocks and the execution time was limited at $1800$ seconds.
\begin{figure}[h!]
\centering
\begin{scriptsize}
\setlength{\tabcolsep}{0.7em}
\setlength{\arrayrulewidth}{.05em}
{\renewcommand{\arraystretch}{1.35}
\begin{tabular}{cccccccccccc}
\hline

$Instance$ & $I1$ & $I2$ & $I3$ & $I4$& $I5$& $I6$& $I7$& $I8$& $I9$& $I10$& $Avrage$ \\ \hline
{$n=$} 32 &1	&1	&1	&0,44	&1	&0.66&	1&	1&	1&	0.60& 0.87\\ \hline
\textcolor[rgb]{1.00,1.00,1.00}{$n=$} 40 &1&1&	1&	1&	1&	0.43&	1&	1&	0.42&	0.50&0.84
\\ \hline
\textcolor[rgb]{1.00,1.00,1.00}{$n=$} 48 &1	&1&	1&	1&	0.40&	1&	0.46&	1&	0.52&	1& 0.88
\\ \hline
\textcolor[rgb]{1.00,1.00,1.00}{$n=$} 56 &1	&1&	0.63&	0.66&	0.59&	1	&1	&0.52&	1&	1&	0.84

 \\ \hline
\textcolor[rgb]{1.00,1.00,1.00}{$n=$} 64 &1&	0.52&	0.56&	1&	0.48&	1&	0.50&	1&	0.64&	1&	0.77

\\ \hline
\textcolor[rgb]{1.00,1.00,1.00}{$n=$} 72 &0.72&	1&	1&	0.67&	1&	0.95&	0.72	&0.43&	1&	0.65&	0.81

 \\ \hline
\textcolor[rgb]{1.00,1.00,1.00}{$n=$} 80 &1&	0.89&	0.63&	1&	0.81&1	&	0.79&	1&	1&0.38	&	0.85
\\ \hline
\textcolor[rgb]{1.00,1.00,1.00}{$n=$} 88 &1&0.65&	0.56&	0.75&	1&	1&	0.58&	1&	0.53&	1&0.81
\\ \hline
\textcolor[rgb]{1.00,1.00,1.00}{$n=$} 96 &1 &	0.52&	1&	1&	0.78	&0.56&	1&	1&	0.46&	1&	0.83
\\ \hline
\end{tabular}
}
\end{scriptsize}\vspace{-.7em}
\capt{3}{Resemblance ratio for $PGA$ Attack. }

\end{figure}\\
Tables 4 and 5 resumes the obtained results regarding the effect of the population size and the mutation probability on the performance of the scheme, adduced in the average execution time and their associated number of generations required for a successful decryption.
\begin{figure}[h!]
\centering
\begin{footnotesize}
\setlength{\tabcolsep}{0.3em}
\setlength{\arrayrulewidth}{.05em}
{\renewcommand{\arraystretch}{1.35}
\begin{tabular}{cccccc:ccccc}
\cline{1-11}
  \multicolumn{1}{r}{} & \multicolumn{5}{c:}{Execution time (ms)} &  \multicolumn{5}{c}{Number of generations} \\
\cline{1-11}
Pop. size &20 & 40 & 80 & 100 & 200 & 20 & 40 & 80 & 100 & 200 \\ \hline
{$n=$} 32&293& 134&540& 1228& 1420& 38& 14&43&50&31\\
\hline
\textcolor[rgb]{1.00,1.00,1.00}{$n=$} 40 &1119& 334&789& 768& 2278& 159& 24&47&28&64\\
\hline
\textcolor[rgb]{1.00,1.00,1.00}{$n=$} 48 &7726&5511&3671&2122&6854&664&497&122&99&49\\
\hline
\textcolor[rgb]{1.00,1.00,1.00}{$n=$} 56 &7230&8478&6775&11282&21717&677&465&433&520&399\\
\hline
\textcolor[rgb]{1.00,1.00,1.00}{$n=$} 64 &\cellcolor{lt}&17692&6489&5128&28328&\cellcolor{lt}&1212&245&196&519\\
\hline
\textcolor[rgb]{1.00,1.00,1.00}{$n=$} 72 &\cellcolor{lt}&75397&64266&26726&29117&\cellcolor{lt}&3206&1603&832&647\\
\hline
\textcolor[rgb]{1.00,1.00,1.00}{$n=$} 80 &\cellcolor{lt}&86263&56250&16937&33820&\cellcolor{lt}&3760&2295&441&343\\
\hline
\textcolor[rgb]{1.00,1.00,1.00}{$n=$} 88 &\cellcolor{lt}&\cellcolor{lt}&150157&80850&110573&\cellcolor{lt}&\cellcolor{lt}&4966&891&1571\\
\hline
\textcolor[rgb]{1.00,1.00,1.00}{$n=$} 96 &\cellcolor{lt}&\cellcolor{lt}&260523&270235&161947&\cellcolor{lt}&\cellcolor{lt}&9483&7991&4655\\
\hline
\textcolor[rgb]{1.00,1.00,1.00}{$n=$} 102 &\cellcolor{lt}&\cellcolor{lt}&\cellcolor{lt}&97101&778023&\cellcolor{lt}&\cellcolor{lt}&\cellcolor{lt}&57194&29132\\
\hline
\textcolor[rgb]{1.00,1.00,1.00}{$n=$} 112 &\cellcolor{lt}&\cellcolor{lt}&\cellcolor{lt}&1883678&1508266&\cellcolor{lt}&\cellcolor{lt}&\cellcolor{lt}&32847&23561\\
\hline
\end{tabular}
}
\end{footnotesize}
\capt{4}{Effect of the population size.}
\end{figure}

\begin{figure}[h!]
\centering
\begin{footnotesize}
\setlength{\tabcolsep}{0.7em}
\setlength{\arrayrulewidth}{.05em}
{\renewcommand{\arraystretch}{1.2}
\begin{tabular}{cccc:ccc}
\cline{1-7}
  \multicolumn{1}{c}{} & \multicolumn{3}{c:}{Execution time (ms)} &  \multicolumn{3}{c}{Nb. of generations} \\
\cline{1-7}
$p_m$ & 0.01 & 0.1 & 0.2 &0.01 & 0.1 & 0.2 \\ \hline
{$n=$} 56 & 2326 &6895 & 9355 & 198& 340&223 \\ \hline
\textcolor[rgb]{1.00,1.00,1.00}{$n=$} 64 &18669 &7592 & 19351& 142 &361& 319\\ \hline
\textcolor[rgb]{1.00,1.00,1.00}{$n=$} 72 &30116 & 24249&35708 &7091 &1153 &362 \\ \hline
\textcolor[rgb]{1.00,1.00,1.00}{$n=$} 80 &27060 & 48024&77951 & 1496&2049 & 928\\ \hline
\textcolor[rgb]{1.00,1.00,1.00}{$n=$} 88 &177154 & 97566&116490 &11739 & 3567&656 \\ \hline
\textcolor[rgb]{1.00,1.00,1.00}{$n=$} 96 &89590 & 180732& 307455& 5486 &3016 & 2681\\ \hline
\textcolor[rgb]{1.00,1.00,1.00}{$n=$} 104 &163024 & 446651&1085037 &9483 &8129 & 11118\\ \hline
\textcolor[rgb]{1.00,1.00,1.00}{$n=$} 112 &1401182 &1269385 &1834316 &28560 &27352 & 33792\\
\hline
\end{tabular}
}
\end{footnotesize}\vspace{-2mm}
\capt{5}{Effect of the mutation probability.}
\end{figure}

Table 6 shows the results related to a conducted experiment aiming to observe the effect of the selection operator. For this purpose, we use $p_s$ the probability that determines the selection operator in each iteration of the algorithm among the roulette wheel selection and the elitist selection, since these operators has a major impact on the evolution of the algorithm, and yet on the quality of the results and its execution time.
\begin{figure}[h!]
\centering
\begin{footnotesize}
\setlength{\tabcolsep}{0.3em}
\setlength{\arrayrulewidth}{.05em}
{\renewcommand{\arraystretch}{1.2}
\begin{tabular}{cccccc:ccccc}
\cline{1-11}
  \multicolumn{1}{c}{} & \multicolumn{5}{c:}{Execution time (ms)} &  \multicolumn{5}{c}{Nb. of generations} \\
\cline{1-11}
$p_s$ & 0& 0.3 & 0.5 &0.8 & 1 & 0 & 0.3 & 0.5 &0.8 & 1 \\
\hline
$n=56$&13356 & 7050& 705 & 2202&1975 &554 &263 & 22&63 &111 \\ \hline
\textcolor[rgb]{1.00,1.00,1.00}{$n=$} 64& 36911& 28108 &41457 & 21429 &31751
 &2396 &1635 &423 & 353& 2060\\ \hline
\textcolor[rgb]{1.00,1.00,1.00}{$n=$} 72& 53259&46138 &28328 & 24080&55448 &2396 &1635 &423 & 353& 2060\\ \hline
\textcolor[rgb]{1.00,1.00,1.00}{$n=$} 80& 71585&62400 &124252 &81368 & 107396& 2023 &2539 &2033 &1797 &2627 \\ \hline
\textcolor[rgb]{1.00,1.00,1.00}{$n=$} 88& 144173&93195 &165704 &127010&247476 &4829 &3651 &6637 & 6051&4316 \\ \hline
\textcolor[rgb]{1.00,1.00,1.00}{$n=$} 96&619685 & 142769&244599 &198579 & 233760&11154 &3404 & 2302 &2025 &1806 \\ \hline
\textcolor[rgb]{1.00,1.00,1.00}{$n=$} 104&793287 & 269490&418359 &280378 &896040 &5669 &3577 & 4186&2609 &9149 \\ \hline
\textcolor[rgb]{1.00,1.00,1.00}{$n=$} 112&3695401 &908263 &1530486 &1408183 & 2202185& 69124&1435 &24821 &28674 & 21263  \\ \hline
\end{tabular}
}
\end{footnotesize}
\vspace{-3mm}
\capt{6}{The impact of $p_s$  variations on the scheme's preference.}
\vspace{-1em}
\end{figure}\newpage
As it is mentioned in Section 4, an improving heuristic has been integrated in the GA search process, therefore, we've conducted an experiment aiming to observe the effect of the heuristic's parameter on the scheme's performance. For this, we fixed the number of chromosomes that goes through the heuristic in each GA iteration to one third of the current population size ($\frac{pop. size}{3}$), where these chromosomes are selected randomly from the current population. The experiment consists of variating the heuristic's number of iterations, the following table shows the obtained results for different values ($nb. iter \in \{ i\times 100| 1\leq i \leq 5\}$). \vspace{-0.3em}
\begin{figure}[h!]
\centering
\begin{footnotesize}
\setlength{\tabcolsep}{0.3em}
\setlength{\arrayrulewidth}{.05em}
{\renewcommand{\arraystretch}{1.2}
\begin{tabular}{cccccc:ccccc}
\cline{1-11}
  \multicolumn{1}{c}{} & \multicolumn{5}{c:}{Execution time (ms)} &  \multicolumn{5}{c}{Nb. of generations} \\
\cline{1-11}
$Nb. iter$ & 100& 200 & 300 &400 & 500 & 100& 200 & 300 &400 & 500  \\
\hline
{$n=$} 40&1016 &971 & 1659& 6873&5419 &142& 107& 172& 496& 274\\
\hline
\textcolor[rgb]{1.00,1.00,1.00}{$n=$} 48&5445 &2684 &840 & 566&1840 & 967&178 & 75& 31&70 \\
\hline
\textcolor[rgb]{1.00,1.00,1.00}{$n=$} 56& 8626& 6306&5171 &6876 &3396 &1393& 582&359 & 312 &65 \\
\hline
\textcolor[rgb]{1.00,1.00,1.00}{$n=$} 64& 13315& 7651&1860 &3118 &6945 & 1958 & 901 & 204 &176 &273 \\
\hline
\textcolor[rgb]{1.00,1.00,1.00}{$n=$} 72&36068 &40622 &35194 &21331 &16239 & 4377&4093 &1581 &1016 & 382\\
\hline
\textcolor[rgb]{1.00,1.00,1.00}{$n=$} 80& 97361& 193045& 267799 &123560 &149724 & 5832 &20946 & 22451 &7637 & 10992\\
\hline
\textcolor[rgb]{1.00,1.00,1.00}{$n=$} 88&179386 & 119915 & 107090&142648 & 203428& 9410 & 15788 & 6761 &10487 & 15331\\
\hline
\textcolor[rgb]{1.00,1.00,1.00}{$n=$} 96& 387153 & 328411 &684342 &239163 &168334 &37303 &24743 &40261 &11541 &18673 \\
\hline
\textcolor[rgb]{1.00,1.00,1.00}{$n=$} 104&929999 & 450031& 529709 &757108   &846461 & 47277 &14494 & 13075& 23874& 20434\\
\hline
\textcolor[rgb]{1.00,1.00,1.00}{$n=$} 112&1833462 &1938746 &797602 &896392 &1412702 &33235 &47957 &46811 &61963 & 42144\\
\hline

\end{tabular}
}
\end{footnotesize}
\vspace{-3mm}
\capt{7}{The effect of the heuristic number of iterations.}
\end{figure}\\
\textbf{\textsc{Comparison with the $\mathit{LLL}$ algorithm. }}
As it is mentioned in the introduction, lattice reduction attack was proposed in \cite{19}, using the $\mathit{LLL}$ algorithm, prsented in \cite{8} by \textit{Lenstra, Lenstra and Lov\'{a}sz}. Analogous to our propsed scheme, this method can also be viewed as a form of heuristic search \cite{14}. 
 We present a comparison with the $LLL$ reduction attack, by running instances (public key $\{b_i\}_{1\leq i\leq n}$ , clear inforamation $m$ and its associated ciphered information $c$), 
 identified by the the paramaters $n, \delta$ and $p$, where, $n$ is the public key size, $\delta$ the density of the knapsack (public key) \cite{16}, and $p$ is the proportion of ones in the clear information $m$;
\begin{align}
\delta = \frac{n}{\log_2 \underset{1\leq i\leq n}\max \,b_i\,}, & \hspace{2em}
 p=\frac{\sum_{i} x_i}{n}\cdot \notag
\end{align}
Furthermore, throughout this expirement, we used two lattice bases, the \textit{Lagarias-Odlyzko} basis \cite{17}, and \textit{Coster, Joux, LaMacchia, Odlyzko, Schnorr, and Stern (CJLOSS)} basis \cite{18}. For this expirementation, we used the $SageMath$  function for $LLL$ algorithm \cite{20}, \cite{15}. Table 8 an 9 present the results of comparison between the $LLL$-based attack and the proposed scheme, obtained by running respectively: $50$ instances per density value and $44$ instances. The aim of the following experiment is to observe the performance of the proposed scheme against the variation public key density.
\begin{footnotesize}
\begin{figure}[H]
\centering
\setlength{\tabcolsep}{0.7em}
\setlength{\arrayrulewidth}{.05em}
{\renewcommand{\arraystretch}{1.}
\begin{tabular}{c:cc}
\hline
$\delta$ & $Proposed \; scheme $ & $LLL \; reduction $ \\
\hline
$ \left[0.6, 0.7\right[$& 0.648 & 0.420 \\
\hline
$ \left[0.7, 0.8\right[$ &0.583 & 0.25\\
\hline
$ \left[0.8, 0.9\right[$ & 0.521&0.271 \\
\hline
$ \left[0.9, 1.0\right[$ & 0.632& 0.183 \\
\hline
\end{tabular}
}
\capt{8}{Success ratios of $LLL\, reduction$ and the proposed scheme.}
\end{figure}
\end{footnotesize}
Throughout the comparison, the $LLL$ algorithm attack has shown lower computational cost than the proposed scheme, which is induced by its differences in data processing approaches and its objectives, which have an important influence on the results quality. In contrast, the success ratio of $LLL$ algorithm is $ 29\%$, while the proposed scheme has a success ratio of $60\%$, attaining a significant difference of $ 31\%$.\\

\textbf{\textsc{Results discussion and comments.}}
\vspace{-2mm}
\begin{itemize}
\item The results in Table 1 shows a minor difference in both average and median decryption time among attacks on one and tow blocks, which, illustrates the efficiency of the proposed parallel approach, even for large instances.

\item Only the parameter's values that manifest consistency has been included in the previous tables, which explains the blank cells in Table~4.
\end{itemize}
\begin{figure}[H]
\begin{small}
\centering
\setlength{\tabcolsep}{0.7em}
\setlength{\arrayrulewidth}{.05em}
{\renewcommand{\arraystretch}{1.1}
\begin{tabular}{ccc|c:c}
\hline
\multicolumn{3}{c|}{Instance} & \multicolumn{2}{c}{Cryptanalysis method} \\
\hline
$n$ & $\delta$ & $p$ & $Proposed\;scheme$ & $\quad LLL \; reduction \quad $\\
\hline
\multirow{4}{*}{32} & 0.627& 0.375&\cmark &\cmark \\
                    &0.711 & 0.343& \cmark &\xmark \\
                    &0.842 & 0.406& \xmark &\xmark  \\
                    &0.914 & 0.250& \cmark &\cmark  \\
\hline
\multirow{3}{*}{40} & 0.634& 0.634& \cmark &\xmark \\
                    &0.769 & 0.525& \cmark &\xmark \\
                    &0.869& 0,425& \cmark  & \cmark  \\
                    &0.930 & 0.500&\cmark &\xmark \\
\hline
\multirow{3}{*}{48} & 0.648& 0,416& \cmark &\cmark \\
                    &0.716 & 0,625& \cmark &\cmark \\
                    &0.800& 0,479& \xmark &\xmark \\
                    &0.906 & 0.542& \cmark & \xmark \\
\hline
\multirow{3}{*}{56} & 0.651& 0,268& \xmark &\xmark \\
                    &0.778 & 0,500 & \xmark &\cmark \\
                    &0.875 & 0,464& \cmark &\xmark \\
                    &0.933 & 0,267& \cmark & \xmark \\
\hline
\multirow{3}{*}{64} &0.680 & 0,250& \cmark &\cmark \\
                    &0.753 & 0.531& \cmark &\xmark \\
                    &0.878 & 0.266& \cmark &\cmark \\
                    &0.941 & 0.438& \xmark & \xmark \\
\hline
\multirow{3}{*}{72} &0.637 & 0.486& \cmark &\xmark \\
                    &0.758 & 0.208& \xmark &\xmark \\
                    &0.878 & 0.472& \cmark &\xmark \\
                    &0.941 & 0.458& \cmark & \xmark \\
\hline
\multirow{3}{*}{80} &0.620 & 0.187& \cmark &\xmark \\
                    &0.721 & 0.412& \xmark &\xmark \\
                    &0.889 & 0.275& \xmark &\xmark \\
                    &0.963 & 0.475& \xmark & \xmark \\

\hline
\multirow{3}{*}{88} &0.651 & 0.443& \xmark &\xmark \\
                    &0.745 & 0.50& \cmark &\xmark \\
                    &0.862 & 0.420& \xmark &\xmark \\
                    &0.977 & 0.454& \cmark & \xmark \\
\hline
\multirow{3}{*}{96} &0.676 & 0.218& \cmark &\xmark \\
                    &0.701 & 0.395& \xmark &\cmark \\
                    &0.857 & 0.427& \cmark &\cmark \\
                    &0.969 & 0.468& \xmark & \xmark \\
\hline

\multirow{3}{*}{104} &0.630 & 0.519& \cmark &\xmark \\
                    &0.717 & 0.298& \cmark &\cmark \\
                    &0.838 & 0.279& \xmark &\xmark \\
                    &0.981 & 0.403& \cmark & \xmark \\
\hline
\multirow{3}{*}{112} &0.674 & 0.473& \cmark &\xmark \\
                    &0.770 & 0.519& \cmark &\cmark \\
                    &0.854 & 0.375& \xmark &\xmark \\
                    &0.971 & 0.384& \cmark & \xmark \\
\hline

\end{tabular}
}
\capt{9}{Comparison between $LLL\, reduction$ and the proposed scheme. \\ \begin{center}
(The symbols \cmark \ and \xmark \ respectively refer to successful and unsuccessful attacks)
\end{center}}\end{small}
\end{figure}
\begin{itemize}
\item The scheme's performance has shown a considerable sensibility towards the parameters $p_m$ and $p_s$ (see Tables 5 and 6). Regarding the selection operator the recommended value as to set the required equilibrium is $p_s\approx0.3$, which gives the elitist selection operator the preponderance; while according to the experimentations, the suitable value for $p_m\approx0.1$.

\item The integrated heuristic has an important role in the global optimization process (decryption), nevertheless, as it is shown in Table 7, an overloaded heuristic can cause delayance in the process of decryption, in other words, the cost of the local optimization can emerge, in case of an encumbered parameters.

\item Regarding that the $LLL$ algorithm is known to be effective against low density knapsack problem, according to the results in Table 8 and 9, the proposed scheme shows no significant sensibility towards the knapsack density $\delta$.
\end{itemize}

\section{Conclusion}

In this paper we present cryptanalysis scheme based on the genetic algorithm, adapted to break the Merkle-Hellman cipher; regarding that in this latter a clear information is ciphered in blocks, the attacking approach is essentially an adapted parallelization  of multiple genetic algorithms, aiming to decrypt a ciphered information in full. Furthermore, the scheme is enhanced with a deliberated cooperation among the search entities (GAs) via the migration operator, although each GA has a distinct target solution (single block clear information), we show that the migration of solutions can be useful under a given condition. Finally, we concluded this paper with some experimental results illustrating the scheme's performance in regards to its parameters.

\end{document}